\documentclass[sigconf]{acmart}
\usepackage[utf8]{inputenc}

\usepackage{caption}
\usepackage{algpseudocode}
\usepackage{amsmath}
\usepackage{amsfonts}
\usepackage{subfigure}
\usepackage{graphicx}

\usepackage{amsfonts}
\usepackage{booktabs}
\usepackage{siunitx}

\usepackage{color}

\usepackage{tikz}

\usepackage{enumitem}

\usepackage{url}
\makeatletter
\def\url@leostyle{%
  \@ifundefined{selectfont}{\def\UrlFont{\sf}}{\def\UrlFont{\small\ttfamily}}}
\makeatother
\newcommand{\eat}[1]{}
\usepackage[linesnumbered,ruled]{algorithm2e}
\usepackage{mdframed} 

\usepackage{xcolor}
\definecolor{light-gray}{gray}{0.9}

\usepackage{amsthm}

\newcolumntype{L}[1]{>{\raggedright\let\newline\\\arraybackslash\hspace{0pt}}m{#1}}

\title{Incentivized Blockchain-based Social Media Platforms: \\A Case Study of \textit{Steemit}}

\author{Chao Li}
\affiliation{%
  \institution{School of Computing and Information \\University of Pittsburgh}
  \city{Pittsburgh}
  \country{USA}
}
\email{chl205@pitt.edu}

\author{Balaji Palanisamy}
\affiliation{%
  \institution{School of Computing and Information \\University of Pittsburgh}
  \city{Pittsburgh}
  \country{USA}
}
\email{bpalan@pitt.edu}

\begin{document}

\begin{abstract}
Advances in Blockchain and distributed ledger technologies are driving the rise of incentivized social media platforms over Blockchains, where no single entity can take control of the information and users can receive cryptocurrency as rewards for creating or curating high-quality contents.
This paper presents an empirical analysis of \textit{Steemit}, a key representative of 
these emerging platforms, to understand and evaluate the actual level of decentralization and the practical effects of cryptocurrency-driven reward system in these modern social media platforms.
Similar to Bitcoin, \textit{Steemit} is operated by a decentralized community, where 21 members are periodically elected to cooperatively operate the platform through the Delegated Proof-of-Stake (DPoS) consensus protocol.
Our study performed on 539 million operations performed by 1.12 million \textit{Steemit} users during the period 2016/03 to 2018/08 reveals that the actual level of decentralization in \textit{Steemit} is far lower than the ideal level, indicating that the DPoS consensus protocol may not be a desirable approach for establishing a highly decentralized social media platform.
In \textit{Steemit}, users create contents as posts which get curated based on votes from other users. 
The platform periodically issues cryptocurrency as rewards to creators and curators of popular posts.
Although such a reward system is originally driven by the desire to incentivize users to contribute to high-quality contents, our analysis of the underlying cryptocurrency transfer network on the blockchain reveals that more than 16\% transfers of cryptocurrency in \textit{Steemit} are sent to curators suspected to be bots and also finds the existence of an underlying supply network for the bots, both suggesting a significant misuse of the current reward system in \textit{Steemit}.
Our study is designed to provide insights on the current state of this emerging blockchain-based social media platform including the effectiveness of its design and the operation of the consensus protocols and the reward system.
\end{abstract}

\maketitle

\section{Introduction}
Recent advances in Blockchain and distributed ledger technologies~\cite{nakamoto2008Bitcoin} are driving the rise of incentivized social media platforms over Blockchains. Examples of such platforms include \textit{Steemit}\footnote[1]{https://steemit.com/}, \textit{Indorse}\footnote[2]{https://indorse.io/}, \textit{Sapien}\footnote[3]{https://beta.sapien.network/} and \textit{SocialX}\footnote[4]{https://socialx.network/}.
Unique features of these blockchain-powered social media platforms include decentralization of data generated in the platforms and deep integration of social platforms with the underlying cryptocurrency transfer networks on the blockchain.
\textit{Steemit} is the first blockchain-powered social media platform that incentivizes both creator of user-generated content and content curators. It has kept its leading position during the last two and a half years and its native cryptocurrency, \textit{STEEM}, has the highest market capitalization among all cryptocurrencies issued by blockchain-based social networking projects. Its market capital is estimated over 266 million USD in 09/30/2018~\cite{Steem_marketcap}.

In \textit{Steemit}, users can create and share contents as blog posts.
Once posted, a blog can get replied, reposted or voted by other users.
Depending on the weight of received votes, posts get ranked and the top ranked posts make them to the front page. 
All data generated by \textit{Steemit} users are stored in the Steem-blockchain~\cite{Steem_blockchain}. Similar to other blockchains like Bitcoin~\cite{nakamoto2008Bitcoin} and Ethereum~\cite{buterin2014next}, data stored in the Steem-blockchain is publicly accessible and it is hard to be manipulated.
At the core of \textit{Steemit} are its decentralization and its cryptocurrency-driven reward system used for rewarding the content creators and curators.
Instead of operating as a single entity like \textit{Reddit} and \textit{Quora}, the \textit{Steemit} platform is operated by a group of 21 witnesses elected by its shareholders (uses owning \textit{Steemit} shares) through the Delegated Proof of Stake (DPoS) consensus protocol~\cite{larimer2014delegated}.
Unlike traditional social media platforms that typically do not reward their users, \textit{Steemit} issues three types of rewards to its users: (1) producer reward; (2) author reward; (3) curation reward. 
The producer rewards are issued to the elected witnesses producing blocks. It incentivizes users of \textit{Steemit} to compete for the top-21 witnesses.
The author rewards and curation rewards are issued to users creating posts (authors) and users voting for posts (curators) respectively. These incentivize authors to produce posts that attract more votes and curators to vote for posts that have higher potential to be voted by other users.
By the end of 2018/08, \textit{Steemit} has issued over 40 million USD worth of rewards to its users~\cite{Steem_reward}.

This paper presents an empirical analysis of \textit{Steemit}, a key representative of emerging blockchain-based incentivized social media platforms.
Our study targets two core features of \textit{Steemit}, namely its decentralized operation and its cryptocurrency-driven reward system. 
By analyzing over 539 million operations performed by 1.12 million users during the period 2016/03 to 2018/08, we aim at obtaining several key insights including the answers for the following set of key questions:

\begin{itemize}[leftmargin=*]
\item Do the members of the witness group in the platform have a high update rate or do the same set of users serve in the group?
What is the power of big shareholders on the decentralization properties of the platform? Is it possible that a single big shareholder can determine who to join the witness group?
\item What are the factors correlated with rewards issued to authors and curators? 
How does the reward system influence users' behavior?
Can incentives be misused by users such as buying votes from bots to promote their posts?

\end{itemize}

Our analysis reveals interesting details on the decentralized operation in \textit{Steemit} and its reward system.
Our study on decentralization in \textit{Steemit} shows that the witness group tends to show a relatively low update rate and its seats may actually be controlled by a few large shareholders. Our analysis also indicates that the majority of top witnesses and top electors form a value-transfer network.
These findings together reveals that the actual level of decentralization in \textit{Steemit} is far lower than the ideal level, indicating that DPoS consensus protocol may not be a desirable approach for establishing a highly decentralized social media platform.
Our analysis of the reward system in \textit{Steemit} shows that author rewards earned by an author are correlated with a number of factors, including the number of followers, number of created posts and owned \textit{Steemit} shares. 
However, our analysis of the \textit{transfer} and \textit{vote} operations show that more than 16\% transfers in the dataset are sent to curators suspected to be bots.
A deeper analysis also reveals the existence of an underlying supply network for the bots, where big shareholders delegate their \textit{Steemit} shares to bots to earn profit.
These results together suggests that the current cryptocurrency-driven reward system in \textit{Steemit} is under substantial misuse that deviates from the original intended goal of rewarding high-quality contents.
We point out that the current consensus protocols and reward systems can hardly achieve their design targets and identify the key reasons and suggest potential solutions.
We believe that the findings in this paper can facilitate the improvement of existing blockchain-based social media platforms and the design of future blockchain-powered websites.

\section{Background}

In this section, we introduce the \textit{Steemit} social media platform that runs over the Steem-blockchain\cite{Steem_blockchain}. 
We present the key use cases of \textit{Steemit} and discuss how \textit{Steemit} leverages the underlying blockchain to function as a decentralized social site that incentivizes  users with cryptocurrency-based rewards.



\noindent \textbf{\textit{Steemit}. }
Users of \textit{Steemit} can create and share contents as blog posts.
A blog post can get replied, reposted or voted by other users.
Based on the weights of received votes, posts get ranked and the top ranked posts make them to the front page. 

\begin{figure}
\centering
{
    \includegraphics[width=8cm,height=6cm]{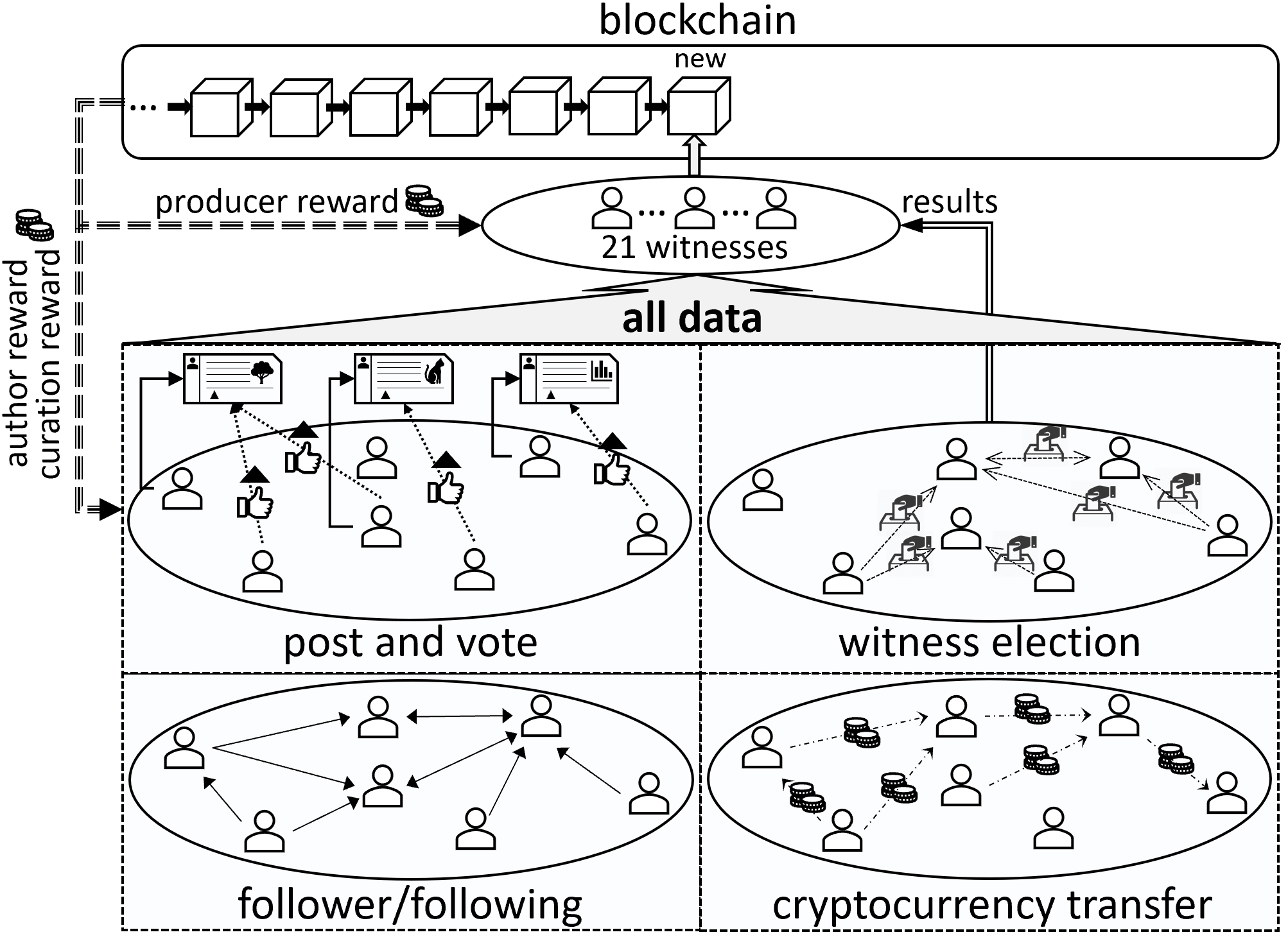}
}
\vspace{-1mm}
\caption {Steem blockchain overview}
\vspace{-6mm}
\label{sec2_1} 
\end{figure}

\noindent \textbf{Steem-blockchain. }
\textit{Steemit} uses the Steem-blockchain\cite{Steem_blockchain} to store the underlying data of the platform as a chain of blocks. Every three seconds, a new block is produced, which includes all confirmed operations performed by users during the last three seconds.
\textit{Steemit} allows its users to perform more than thirty different type of operations. In Figure~\ref{sec2_1}, we display four categories of operations that are most relevant to the analysis presented in this paper. 
While post/vote and follower/following are common features offered by social sites (e.g., Reddit~\cite{stoddard2015popularity} and Quora~\cite{wang2013wisdom}), operations such as witness election and cryptocurrency transfer are features specific to blockchains.

\noindent \textbf{Witness election and DPoS. }
Witnesses in \textit{Steemit} are producers of blocks, who continuously collect data from the entire network, bundle data into blocks and append the blocks to the Steem-blockchain.
The role of witnesses in \textit{Steemit} is similar to that of miners in Bitcoin~\cite{nakamoto2008Bitcoin}. 
In Bitcoin, miners keep solving Proof-of-Work (PoW) problems and winners have the right to produce blocks.
With PoW, Bitcoin achieves a maximum throughput of 7 transactions/sec ~\cite{croman2016scaling}. However, transaction rates of typical mainstream social sites are substantially higher. For example, Twitter has an average throughput of more than 5000 tweets/sec~\cite{Twitter}.
Hence, \textit{Steemit} adopts the Delegated Proof of Stake (DPoS)~\cite{larimer2014delegated} consensus protocol to increase the speed and scalability of the platform without compromising the decentralized reward system of the blockchain.
In DPoS systems, users vote to elect a number of witnesses as their delegates. In \textit{Steemit}, each user can vote for at most 30 witnesses. The top-20 elected witnesses and a seat randomly assigned out of the top-20 witnesses produce the blocks.
With DPoS, consensus only needs to be reached among the  21-member witness group, rather than the entire blockchain network like Bitcoin. This significantly improves the system throughput.


\noindent \textbf{Cryptocurrency - shares and rewards. }
In \textit{Steemit}, each vote casting a post or electing a witness is associated with a weight that is proportional to the shares of \textit{Steemit} held by the voter.
Like most blockchains, the Steem-blockchain issues its native cryptocurrencies called \textit{STEEM} and Steem Dollars (\textit{SBD}).
To receive shares of \textit{Steemit}, a user needs to `lock' \textit{STEEM}/\textit{SBD} in \textit{Steemit} to receive Steem Power (\textit{SP}) at the rate of $1\ STEEM = 1\ SP$ and each $SP$ is assigned about 2000 vested shares (\textit{VESTS}) of \textit{Steemit}.
A user may withdraw invested \textit{STEEM}/\textit{SBD} at any time, but the claimed fund will be automatically split into thirteen equal portions to be withdrawn in the next thirteen subsequent weeks.
For example, in day 1, Alice may invest 13 \textit{STEEM} to \textit{Steemit} that makes her vote obtain a weight of 13 \textit{SP} (about 26000 \textit{VESTS}). Later, in day 8, Alice may decide to withdraw her 13 invested \textit{STEEM}. Here, instead of seeing her 13 \textit{STEEM} in wallet immediately, her \textit{STEEM} balance will increase by 1 \textit{STEEM} each week from day 8 and during that period, her \textit{SP} will decrease by 1 \textit{SP} every week.

Through Steem-blockchain, \textit{Steemit} issues three types of rewards to its users: (1) producer reward; (2) author reward and (3) curation reward.
The amount of producer reward is about 0.2 \textit{STEEM} per block in 2018/08, meaning that a witness producing blocks for a whole day can earn about 14,400 \textit{STEEM}.
Each day, the Steem-blockchain issues a number of \textit{STEEM} (about 53,800 \textit{STEEM} per day in 2018/08) to form the post reward pool and posts compete against each other to divide up the reward pool based on the total weight of votes received within seven days from the post creation time. Here, 75\% of reward received by a post goes to the post author and the rest is shared by the post curators based on their vote weight. 

In the rest of this paper, for ease of exposition and comparison, we transfer all values of \textit{STEEM/SBD/SP/VESTS} to US dollars (\textit{USD} denoted by \$), based on their median transfer rate from 2018/01 to 2018/08, which is $\$1 = 1\ SBD \approx 0.4\ STEEM = 0.4\ SP \approx 800\ VESTS$~\cite{SteemPrice}.

\section{Dataset and preliminary results}
In this section, we describe our data collection methodology and present some preliminary results including the growth of \textit{Steemit} over time and the usage of operations in the platform. 

\subsection{Data collection}
\label{sec3.1}
The Steem-blockchain offers an Interactive Application Programming Interface (API) for developers and researchers to collect and parse the blockchain data~\cite{SteemAPI}.
From block 1 (created at 2016/03/24 16:05:00) to block 25,563,499 (created at 2018/09/01 00:00:00), we collected 539,817,204 operations performed by 1,120,166 users of \textit{Steemit} during the period 2016/03 to 2018/08. The collected data also includes 121,619,828 virtual operations performed in the \textit{Steemit} platform, such as issuing rewards to witnesses, authors and curators, during the same time period.
In the data collected, we recognized 36 types of operations performed by users and 11 types of virtual operations. In table~\ref{t1}, we summarize the key operations (OP) and virtual operations (VO) focused in this paper.
We have recently released a dataset for \textit{Steemit} named \textit{SteemOps} at: \\ 
\centerline{\url{https://github.com/archerlclclc/SteemOps}}

\begin{table}
\small
\centering
\begin{tabular}{|p{2.7cm} |p{5.0cm}|} \toprule 
{\textbf{OP (social)}} & {\textbf{Description}} 
\\ \midrule
    comment & users create posts, reply to posts or replies \\
    vote & users vote for posts \\
    custom\_json & users follow other users, repost a blog \\ 
    \midrule
{\textbf{OP (witness-election)}} & {\textbf{Description}} 
\\ \midrule
    witness\_update & users join the witness pool to be elected, witnesses in pool update their information \\
    witness\_vote & users vote for witnesses by themselves \\
    witness\_proxy & users cast votes to the same witnesses voted by another user by setting that user as their election proxy \\
    \midrule
{\textbf{OP (value-transfer)}} & {\textbf{Description}} 
\\ \midrule
    transfer & users transfer \textit{STEEM}/\textit{SBD} to other users \\
    transfer\_to\_vesting & users transfer \textit{STEEM}/\textit{SBD} to \textit{VESTS} \\
    delegate\_vesting\_shares & users delegate \textit{VESTS} to other users \\ 
    withdraw\_vesting & users transfer \textit{VESTS} to \textit{STEEM} \\ 
    \midrule
{\textbf{VO (reward)}} & {\textbf{Description}} 
\\ \midrule
    producer\_reward & platform issues rewards to producers \\
    author\_reward & platform issues rewards to authors \\
    curation\_reward & platform issues rewards to curators \\ 
    \bottomrule
\end{tabular}
\caption {\small Summary of operations and virtual operations}
\vspace{-10mm}
\label{t1}
\end{table}

\subsection{Preliminary results}

\begin{figure*}
\minipage{0.32\textwidth}
  \includegraphics[width=\columnwidth]{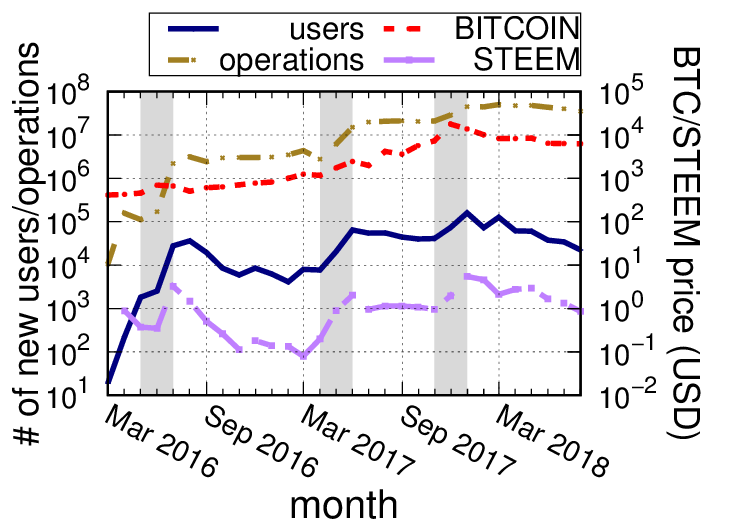}
  \caption {\small New users/operations per month (2016/03 to 2018/08)}
  \vspace{-2mm}
  \label{sec3_1}
\endminipage\hfill
\minipage{0.32\textwidth}
  \includegraphics[width=\columnwidth]{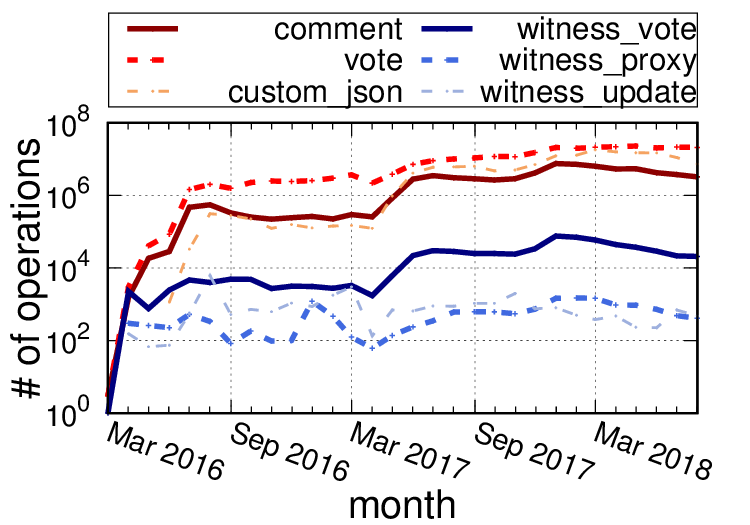}
  \caption {\small New social/witness-election operations per month (2016/03 to 2018/08)}
  \vspace{-2mm}
  \label{sec3_2}
\endminipage\hfill
\minipage{0.32\textwidth}%
  \includegraphics[width=\columnwidth]{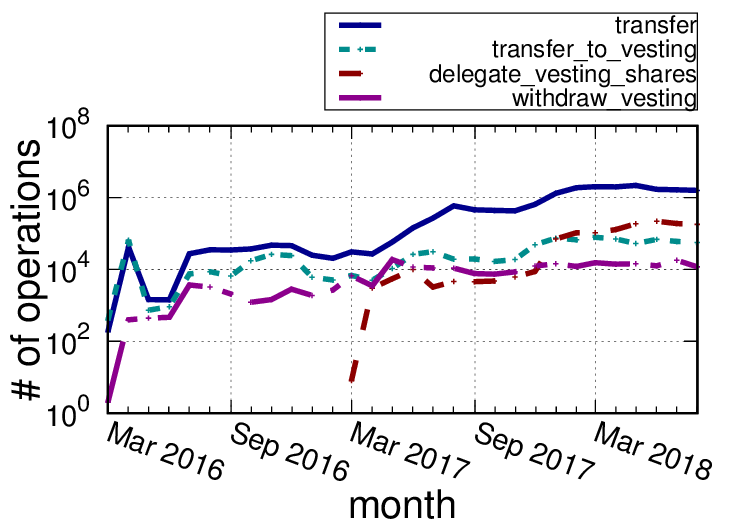}
  \caption {\small New value-transfer operations per month (2016/03 to 2018/08)}
  \vspace{-2mm}
  \label{sec3_3}
\endminipage
\end{figure*}

\noindent \textbf{Growth of \textit{Steemit}. }
We first investigate the growth of \textit{Steemit} over time and find that the platform growth is highly impacted by the cryptocurrency market.
In Figure~\ref{sec3_1}, we plot the numbers of newly registered users and newly performed operations per month and the changes of \textit{Bitcoin} and \textit{STEEM} price during the period 2016/03 to 2018/08. 
Apart from the initial boost in the first month, the platform witnessed three times of robust growth, which happened during 2016/05 to 2016/07, 2017/04 to 2017/06 and 2017/11 to 2018/01, respectively.
We can observe a strong correlation between the monthly increment of users and the changes in the \textit{STEEM} price.
This is due to the fact that more \textit{Steemit} users investing in trading may drive up the \textit{STEEM} price while higher \textit{STEEM} price may in turn attract more people joining \textit{Steemit}.
Next, in cryptocurrency market, the changes of \textit{Bitcoin} price are usually seen as the most important market signals.
During all the three rising periods, we see the surge in \textit{Bitcoin} price, which illustrates the high influence of the cryptocurrency market on the growth of \textit{Steemit} and also suggests that most \textit{Steemit} users may have a background understanding of cryptocurrency and blockchain.
Finally, by comparing the user curve and the operation curve, we find that even though the user growth rate drops after all the three rising periods, the operation growth rate keeps maintaining stability after boosting, which may reflect that most users who joined \textit{Steemit} during the rising periods remain active after the end of the rising periods.

\noindent \textbf{Usage of operations. }
As discussed in section~\ref{sec3.1}, users in \textit{Steemit} may perform a variety of operations and these operations are recorded by the Steem-blockchain. By investigating the usage of these operations, we aim to answer the following questions: 
(1) which categories of operations are more frequently performed by users?, (2) do users use \textit{Steemit} more like a social media platform or like a cryptocurrency wallet?
In Figure~\ref{sec3_2} and Figure~\ref{sec3_3}, we plot the numbers of social/witness-election operations and value-transfer operations performed in different months, respectively.
The functionality of these operations is described in Table~\ref{t1}.
Among the three categories of operations, the social operations show the highest utilization rate, which indicates that users are using more social functions offered by \textit{Steemit} than transfer functions.
Among the three social operations, the \textit{vote} operation is the most frequently used one. In 2018/08, more than 21 million votes were cast to posts. Unlike other voting-based social media platforms such as \textit{Reddit}, in \textit{Steemit}, votes cast by users owning \textit{Steemit} shares have real monetary value, which may incentivize \textit{Steemit} users to keep voting for posts with some frequency to avoid wasting their voting power. We will discuss more about voting and rewarding mechanisms in section~\ref{s5}.
Among the four value-transfer operations, users perform the \textit{transfer} operation more frequently.
Since the \textit{transfer} operation is the only operation among the four that is not associated with \textit{VESTS}, namely shares of \textit{Steemit}, the fact may reflect that more trading behaviors are happening in \textit{Steemit} than investing behaviors.
Finally, the number of performed witness-election operations are relatively small, compared with the other two categories. 
Each month, we see only thousands of users setting or updating their witness votes through the \textit{witness\_vote} and \textit{witness\_proxy} operations and only hundreds of users joining the witness pool or updating their witness information through the \textit{witness\_update} operation.
This reflects a relatively low participation rate in witness election regarding both elector and electee and could impact the level of decentralization in the platform. We discuss witness election in more detail in section~\ref{s4}.

\section{Decentralization in \textit{Steemit}}
\label{s4}
As a blockchain-based social media platform, \textit{Steemit} distinguishes itself from traditional social sites through the decentralization brought by the blockchain.
In this section, we investigate the actual level of decentralization in \textit{Steemit} by analyzing the witness election process of DPoS in detail.

\subsection{Decentralized platform operation}

\noindent \textbf{Centralization and decentralization. }
In traditional social media platforms, such as \textit{Reddit} and \textit{Quora}, a single entity (i.e., Reddit, Inc. and Quora, Inc.) owns the complete data generated by users and operates the websites.
In contrast, \textit{Steemit} not only open sources its front-end, \textit{Condenser} and the back-end Steem-blockchain~\cite{SteemOpenSource}, but also makes all its data in the blockchain available for public access~\cite{Steem_blockchain}.
Rather than functioning as a single entity, the \textit{Steemit} platform is operated by a group of 21 witnesses elected by its users.
Any user in \textit{Steemit} can run a server, install the Steem-blockchain and synchronize the blockchain data to the latest block.
Then, by sending a \textit{witness\_update} operation to the network, the user can become a witness and have a chance to operate the website and earn producer rewards if he or she can gather enough support from the electors to join the 21-member witness group.

\noindent \textbf{\textit{Steemit-2} and \textit{STEEM-2}. }
As anyone can copy the code and data of \textit{Steemit}, one may naturally doubt that an adversary, say Bob, may build a `fake' \textit{Steemit} platform, \textit{Steemit-2} that has exactly same the functionality and historical data as \textit{Steemit}.
To distinguish \textit{Steemit-2} from \textit{Steemit}, we name the cryptocurrency issued by \textit{Steemit-2} as \textit{STEEM-2} (and also \textit{SBD-2}).
Here, a natural question that arises is what makes people believe that \textit{Steemit}, rather than \textit{Steemit-2}, is the `real' one?
In the decentralized network, the opinion of `which one is real' is determined by the consensus of \textit{Steemit} users or, to be more precise, the shareholders.
With the DPoS consensus protocol, each block storing data of \textit{Steemit} is signed by a top witness elected by the shareholders, which may represent the consensus among the shareholders.
Therefore, unless most of shareholders switch to \textit{Steemit-2}, the new blocks generated in \textit{Steemit-2} will be signed by witnesses elected by a few shareholders and will not be recognized by the entire community.

\noindent \textbf{Factors affecting decentralization of \textit{Steemit}. }
In our work, we study the characteristics of decentralization in \textit{Steemit} by analyzing the witness election process. 
In general, we consider \textit{Steemit} to have an ideal level of decentralization if members of the 21-member witness group are frequently updated and if these members all have different interests.
We also consider \textit{Steemit} to have a relatively high decentralization closer to the ideal level if it allows more people to join the 21-member witness group, if the power of big shareholders is not decisive and if the election is not heavily correlated with value-transfer operations. 
We investigate these aspects in the following subsection.

\subsection{Analyzing witness election}
\label{s4_2}

\begin{figure}
\centering
{
    \includegraphics[width=7cm,height=6cm]{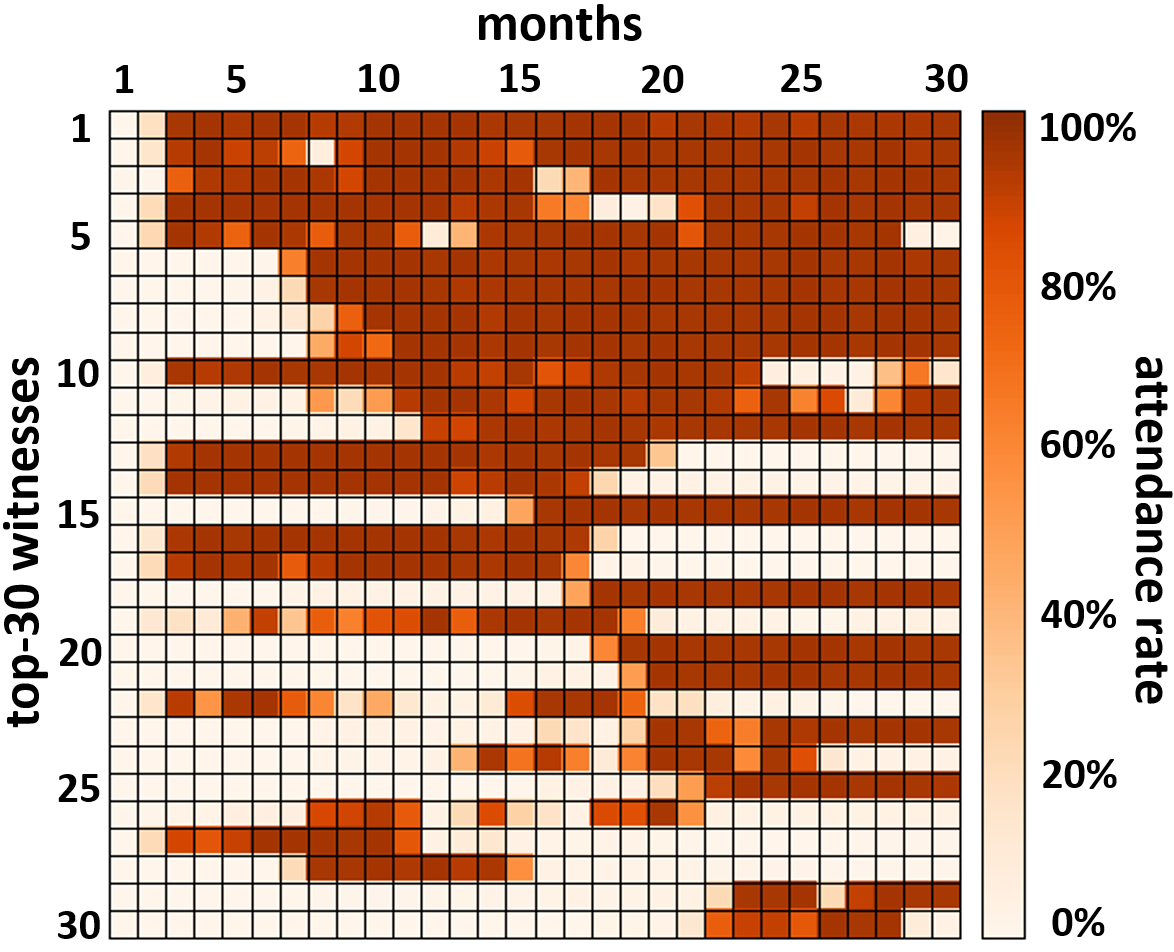}
}
\caption {\small Heatmap of top-30 witnesses' attendance rate in the 21-member witness group during 30 months from 2016/03 to 2018/08}
\vspace{-3mm}
\label{sec4_1} 
\end{figure}

\noindent \textbf{Update of the 21-member witness group. }
To investigate the update of the 21-member witness group, we extract the producer of each block from block 1 to block 25,563,499 and plot the results as a heatmap in Figure~\ref{sec4_1}.
We first compute the number of blocks produced by each witness and sort the witnesses based on their produced blocks in total. For the top-30 witnesses that have the highest number of produced blocks, we plot their attendance rate in the 21-member witness group during the thirty months.
During a month that has thirty days, there should be $30*24*60*60/3=$ 864,000 blocks generated because blocks are generated every three seconds.
For every 21 blocks (63 seconds), the 21 elected witnesses are shuffled to determine their order for generating the next 21 blocks.
Therefore, if a witness has a 100\% attendance rate in the elected group, it can at most produce $864000/21 \approx$ 41,142 blocks in a thirty-day month.
In Figure~\ref{sec4_1}, we find that most of the top-30 witnesses, once entering the 21-member witness group, maintained a high attendance rate closer to 100\% for a long time.
From month 12 to month 30, namely one and a half year, the top-12 witnesses firmly held at least 10 seats, nearly half of positions in the 21-member witness group.
From month 23 to month 30, 17 seats were held by 18 witnesses.
From witness 13 to witness 30, we can observe a transition period, namely month 15 to month 20, during which the places of nine old witnesses were gradually taken by nine new witnesses.
Overall, the 21-member witness group tends to show a relatively low update rate.
The majority of seats were firmly controlled by a small group of witnesses.
We do observe some switch of seats but that happened only in a low frequency.

\begin{figure}
\centering
{
    \includegraphics[width=6cm,height=5.7cm]{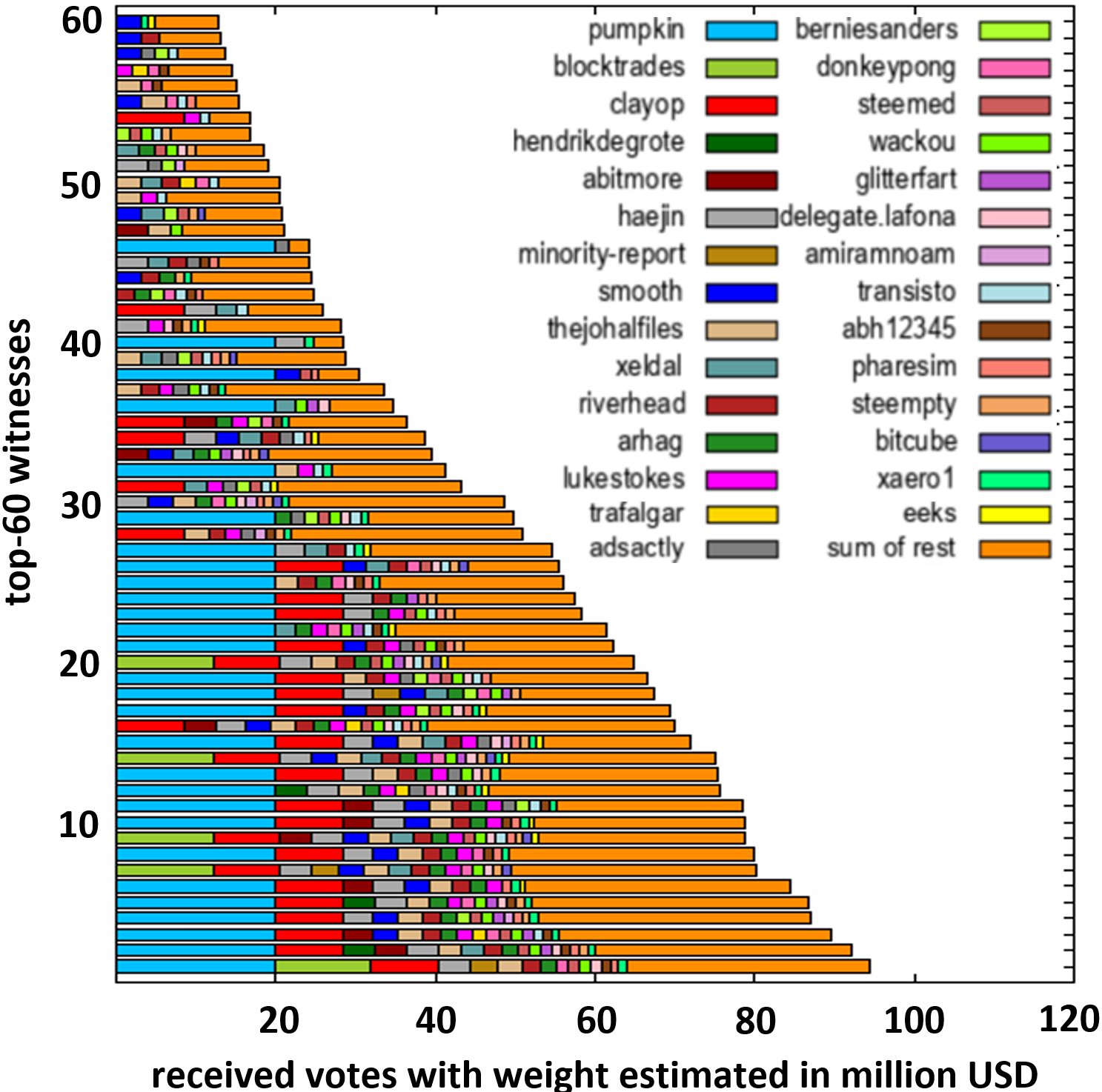}
}
\caption {\small A snapshot of weighted votes received by top-60 witnesses at block 25,563,499. The votes are weighted by shares (estimated in \textit{USD}) owned by the electors.
We show the votes cast by the top-29 electors owning the highest weight and use `sum of rest' to represent the sum of weighted votes cast by all other electors. }
\vspace{-3mm}
\label{sec4_2} 
\end{figure}

\noindent \textbf{Power of big shareholders. }
Next, we investigate the influence of big shareholders in the witness election.
As described in Table~\ref{t1}, a user has two ways to vote for witnesses. 
The first option is to perform \textit{witness\_vote} operations to directly vote for at most 30 witnesses. 
The second option is to perform a \textit{witness\_proxy} operation to set another user as an election proxy.
For example, Alice may set Bob to be her proxy. Then, if both Alice and Bob own \$100 worth of shares, any vote cast by Bob will be associated with a weight of \$200 worth of shares. Once Alice deletes the proxy, the weight of Bob's votes will reduce to \$100 worth of shares immediately.
In Figure~\ref{sec4_2}, we plot the stacked bar chart representing a snapshot of weighted votes received by the top-60 witnesses, who have produced the highest amounts of blocks, at block 25,563,499.
The figure shows the distribution of votes cast by the top-29 electors whose votes have the highest weight, either brought by their own shares or shares belonging to users setting these electors as proxy.
The sum of weighted votes cast by all other electors outside the top-29 is represented as `sum of rest'.
From Figure~\ref{sec4_2}, we see a few top electors have their votes weighted by striking amounts of shares.
The top-1 elector has his or her votes weighted by \$19,800,000 worth of shares. A deeper investigation regarding the top-1 elector shows that all the shares affecting his or her weight are not directly owned by this user, but owned by another user, who set the top-1 elector as proxy.
The runner-up elector, which is the account belonging to the main exchange used by \textit{Steemit} users, has its votes associated with a weight of \$12,100,000 worth of shares.
From the figure, we see all the 27 witnesses voted by the top-1 elector enter the top-50, all the 18 witnesses receiving votes from both the top-2 electors enter the top-30 and the only witness receiving votes from all the top-3 electors becomes the top-1 witness.
We can also observe that 19 out of the top-20 witnesses receive at least two votes from the top-3 electors and 29 out of the top-30 witnesses receive at least one vote from the top-3 electors.
As illustrated by the results, the distribution of weight of votes in witness election is heavily skewed, which suggests that the election of 21 witnesses may be significantly impacted by a few big shareholders, 
This phenomenon may not be a good indication for a decentralized social media platform. In the worst case, if the 21-member witness group is controlled by a single shareholder, the platform will simply function as a centralized model.

\begin{figure}
\centering
{
    \includegraphics[width=7cm,height=6cm]{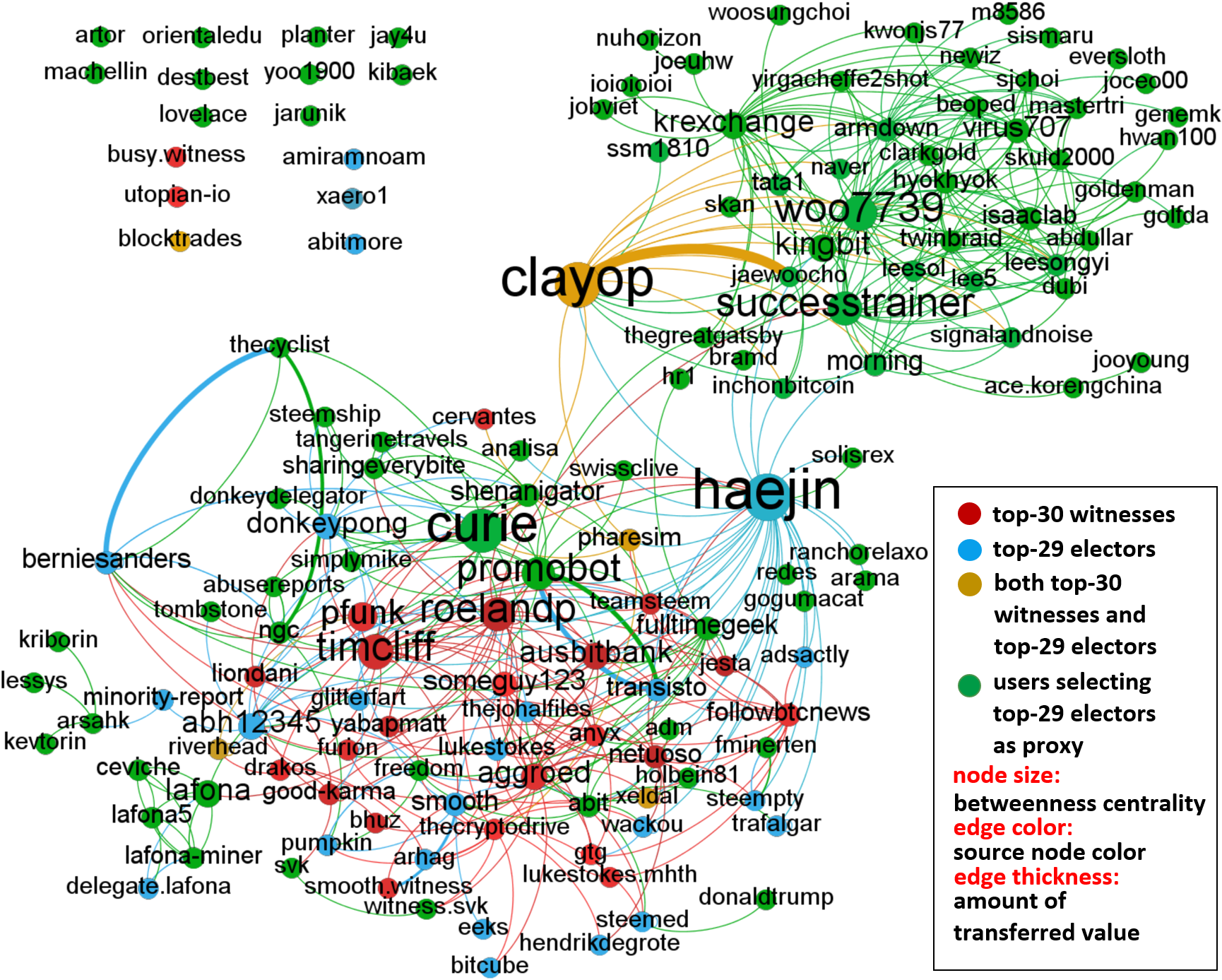}
}
\caption {\small Graph of value-transfer operations performed among the top-30 witnesses, the top-29 electors and the accounts selecting these electors as proxy. The graph was plotted using Gephi~\cite{bastian2009gephi}.}
\label{sec4_3} 
\vspace{-5mm}
\end{figure}

\noindent \textbf{Value-transfer operations among election stakeholders. }
Finally, we investigate the value-transfer operations performed among the top-30 witnesses, top-29 electors and the accounts selecting top-29 electors as proxy.
The data is plotted as a directed graph in Figure~\ref{sec4_3}, where edges are colored by their source node color.
The edge thickness represents the amount of transferred value from source to target, which is the sum of value transferred through \textit{transfer} and \textit{transfer\_to\_vesting} operations.
Since most \textit{Steemit} users use runner-up elector to trade cryptocurrency, the graph does not show edges connected with that account.
Our first observation about this graph is that only two out of top-30 witnesses and three out of top-29 electors never perform value-transfer operations while all other investigated users form a value-transfer network, which has 3.34 average degree, 0.21 average clustering coefficient and 3.96 average path length.
In the network, we find a cluster of users who select top-29 electors as proxy. After manually checking the profiles of these users, we find that this cluster represents a community of Korean users, which is connected to the rest of the network mainly through several leaders of the Korean community.
Overall, what we observed from the value-transfer operations suggests that the majority of the investigated election stakeholders have economic interactions, which may not be a good indication of a perfectly decentralized witness group where the members are expected to have different interests.

\subsection{Discussion}

Our study in this section demonstrates that the 21-member witness group tends to show a relatively low update rate and these seats may actually be controlled by a few big shareholders.
Our study also indicates that the majority of top witnesses and top electors form a value-transfer network and have economic interactions.
Together, these results suggest that the actual level of decentralization in \textit{Steemit} is far lower than the ideal level.
One key reason for the low decentralization is the use of DPoS consensus protocol. 
The DPoS consensus protocol has been widely adopted by mainstream blockchain-based platforms such as BTS and EOS and has been proved to be an effective approach of enhancing transaction rates of blockchains. 
However, there has been a long debate surrounding decentralization of DPoS.
The opponents of DPoS censure that DPoS-powered platforms trade decentralization for scalability as the consensus in these platforms is only reached among a small committee (e.g., the 21-member witness group in \textit{Steemit}), instead of among all interested members (e.g., all miners in Bitcoin and Ethereum powered by Proof-of-Work (PoW) consensus).
The supporters of DPoS argue that those PoW-powered platforms has been controlled by a few mining pools, showing even lower decentralization than the DPoS-powered platforms.
The data-driven analysis in this section deeply investigates the underlying behaviors of participants interested in the core component of DPoS, namely the witness election.
The results reveal that the current electoral system is making the decentralization quite fragile, as the committee intended to be decentralized is actually quite centralized in practice.
A quick solution to address the symptoms is to restrict the power of big stakeholders, such as cutting the number of times that big stakeholders can vote in the election.
A better way of solving the problem is to replace DPoS with more advanced consensus protocols, that can form the committee without an election involving interactions among users.
For example, Algorand~\cite{gilad2017algorand}, a recent cryptocurrency, proposed a new Byzantine Agreement (BA) protocol that makes the election be fairly performed by the Verifiable Random Functions in cryptography, rather than by users.


\section{reward system in \textit{Steemit}}
\label{s5}
A core feature of \textit{Steemit} is its reward system. 
As a blockchain-based social media network that issues its own cryptocurrencies, \textit{Steemit} leverages its native cryptocurrencies to incentivize authors and curators of posts for rewarding contributors to the platform, who either create good contents or screen out good contents.
In this subsection, we theoretically analyze the reward system in \textit{Steemit} and study the factors correlated with rewards earned by authors and curators.
We also jointly investigate the value-transfer operations and \textit{vote} operations to understand if the reward system is being misused by \textit{Steemit} users, such as buying votes from bots to promote their posts for the purpose of earning higher profit. Such misuse may deviate from the original intended goal of the \textit{Steemit} reward system.


\subsection{Reward system}
\label{s5_1}
To the best of our knowledge, \textit{Steemit}~\cite{Steem_Whitepaper} never formally published the details of its reward system, so we investigated its source code to understand its reward system~\cite{SteemOpenSource} in detail.
Each time a user votes for a post, the vote contributes a certain amount of reward shares (\textit{rshares}, denoted as $rs$) to that post. It is computed as
$$ rs = e\_VESTS*\frac{vp*vw-0.0049}{50} $$
where $e\_VESTS$ refers to effective vesting shares (\textit{VESTS}), $vp$ stands for voting power and $vw$ denotes voting weight.
In \textit{Steemit}, users can deposit or withdraw \textit{VESTS} through \textit{transfer\_to\_vesting} operation and \textit{withdraw\_vesting} operation, respectively. However, as described in Table~\ref{t1}, a user may also delegate \textit{VESTS} to other users through \textit{delegate\_vesting\_shares} operation.
For example, if Alice owns 100 \textit{VESTS} and if she delegates 50 \textit{VESTS} to Bob and if she also receives 30 \textit{VESTS} delegated to her by Carol, then her $e\_VESTS=100-50+30=80\ VESTS$.
A user may set voting weight $vw$ to any value between 0\% and 100\%.
\textit{Steemit} leverages voting power $vp$ to restrict the number of weighted votes cast by users per day.
Initially, each user has $vp=100\%$.
Then, if a user casts a vote to a post, this user's $vp$ will drop to $(1-\frac{1-0.0049}{50})*100\%$, which is roughly 98\% if he/she sets $vw=100\%$.
That is, if a user keeps voting, his/her $vp$ will also keep dropping. Once $vp$ drops to 0\%, his/her vote will contribute no $rshares$ to posts and the user and post authors will not earn rewards from this vote.
Each day, $vp$ recovers 20\% before it is back to 100\%.

\begin{figure*}
\centering
\subfigure[{\small Followers}]
{
   \label{sec5_1_1}
  \includegraphics[width=0.66\columnwidth]{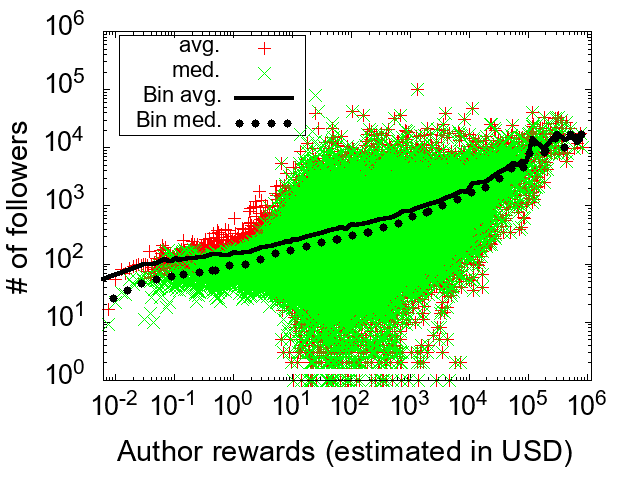}
}
\subfigure[{\small Post numbers}]
{
	\label{sec5_1_2}
    \includegraphics[width=0.66\columnwidth]{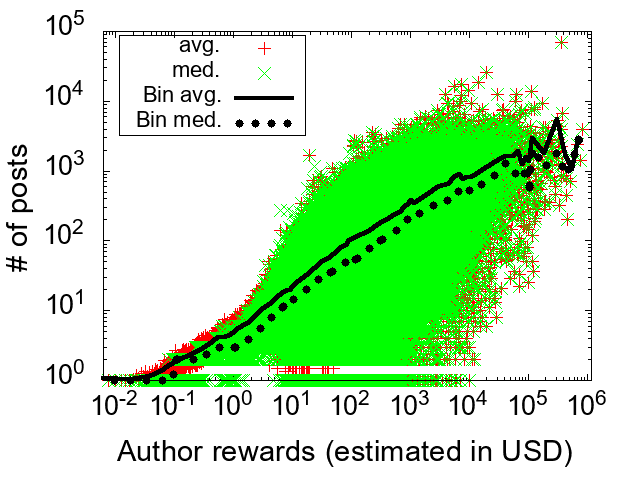}
}
\subfigure[{\small Vesting shares}]
{
   \label{sec5_1_3}
  \includegraphics[width=0.66\columnwidth]{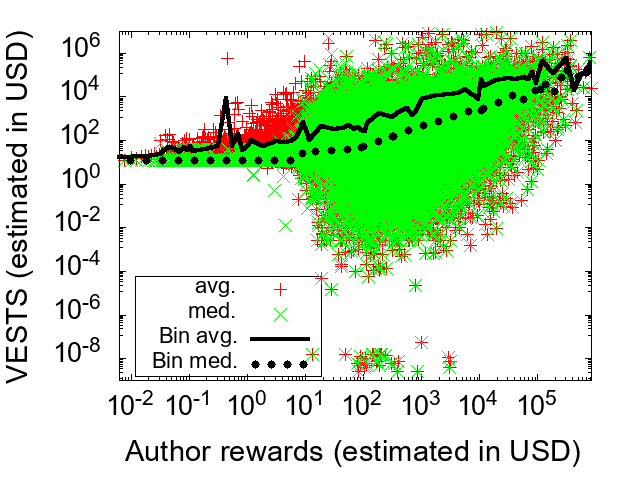}
}
\caption{Scatter plot investigating correlation between followers (post numbers, vesting shares) and author rewards}
\vspace{-2mm}
\label{sec5_1}
\end{figure*}

\begin{figure*}
\minipage{0.32\textwidth}
  \includegraphics[width=\columnwidth]{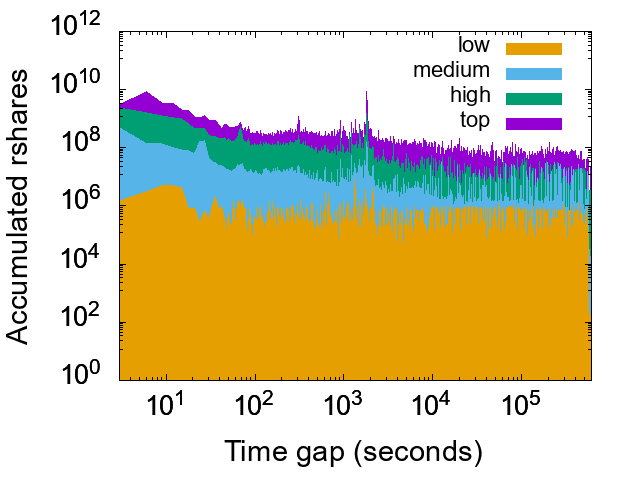}
  \caption {\small Distribution of accumulated \textit{rshares} over time after post creation time}
  \vspace{-2mm}
  \label{sec5_2}
\endminipage\hfill
\minipage{0.32\textwidth}
  \includegraphics[width=\columnwidth]{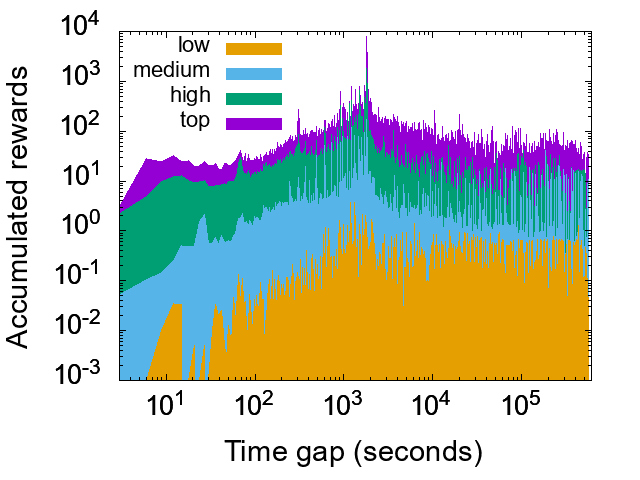}
  \caption {\small Distribution of accumulated curation rewards over time}
  \vspace{-2mm}
  \label{sec5_3}
\endminipage\hfill
\minipage{0.32\textwidth}%
  \includegraphics[width=\columnwidth]{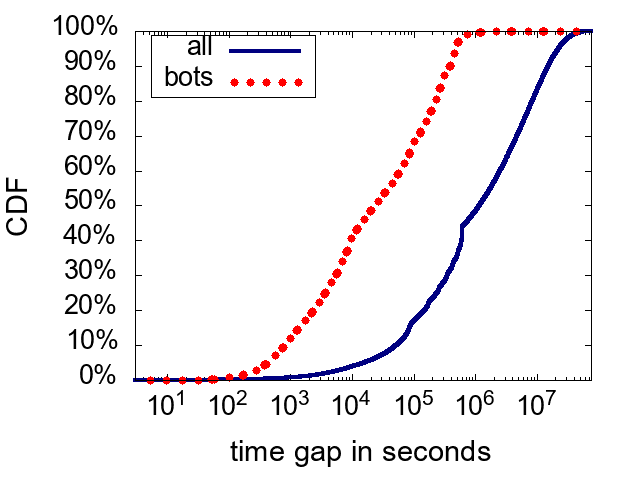}
  \caption {\small CDF investigating the time gap between post creation time and bot voting time}
  \vspace{-2mm}
  \label{sec5_4}
\endminipage
\end{figure*}

A post, after being created, can accumulate \textit{rshares} from received votes during a 7-day time window. At the end of the time window, the post will use the accumulated \textit{rshares} to compete with accumulated \textit{rshares} of other posts to divide up the post reward pool (about 53,800 \textit{STEEM} per day in 2018/08).
After that, 75\% of the post reward (denoted as $pr$) is directly issued to the post author as author reward (denoted as $ar$) and the rest 25\% is finally shared by all curators who voted for the post during the 7-day time window, namely their curation rewards (denoted as $cr$).
The curation reward received by a single curator can be computed as:
$$ cr = 0.25*pr*\frac{\sqrt{rs_{b}+rs}-\sqrt{rs_{b}}}{\sqrt{rs_{T}}}*min(\frac{td}{30},1) $$
As can be seen, $cr$ is computed from 25\% of $pr$, but is affected by two factors. 
The first factor is the ratio between $\sqrt{rs_{b}+rs}-\sqrt{rs_{b}}$ and $\sqrt{rs_{T}}$, where $rs_{b}$ denotes the $rshares$ accumulated by the post before this user's vote and $rs_{T}$ refers to the total accumulated $rshares$ by the post during the 7-day time window.
This is a very interesting factor as it suggests that curators who want to earn higher $cr$ should vote for posts as early as possible to make $rs_{b}$ smaller. It also suggests that such curators vote for posts that have higher probability to accumulate higher $rs_{T}$ by the end of 7-day time window.
However, the second factor forces curators to reconsider the early vote strategy. 
It is the minimum of $\frac{td}{30}$ and 1, where $td$ denotes the time difference in minutes between the post creation time and voting time. For example, if Alice votes for a post one minute after the post creation time, she will only receive $\frac{1}{30}$ of the curation rewards assigned to her while the rest $\frac{29}{30}$ is again issued to the post author. 
Instead, if Alice casts her vote 30 minutes after the post creation time, she can completely earn the assigned curation rewards.
Because of the second factor, a post author usually receives more than $75\%$ of the post reward.
In summary, the two factors in the curation reward equation make it hard to determine the best strategy of the voting time in the reward earning game. However, it is clear that voting for posts that are likely to be voted by more users in the future is certainly helpful.

\subsection{Factors correlated with rewards}
After theoretically analyzing the reward system, we now investigate the blockchain data to learn the factors correlated with author rewards and curation rewards.

\noindent \textbf{Author rewards. }
Regarding author rewards, we investigate three factors: (1) number of followers owned by an author; (2) number of posts created by an author; (3) amount of vesting shares (\textit{VESTS}) owned by an author.
In order to gauge the correlation between these factors and author rewards, for each factor, we plot in Figure~\ref{sec5_1} the average and median of values of the factor ($y$) against author rewards estimated in USD ($x$).
Following the methodology used by Kwak \textit{et al.}~\cite{kwak2010twitter}, we also bin the value of author rewards in log scale and plot the average and median per bin in lines.
In Figure~\ref{sec5_1_1}, we see the line of average is always above the line of median, indicating that some users with followers more than average do not receive expected degree of support from their followers regarding their posts. The gap between the two line is larger for users receiving lower author rewards, but becomes relatively small for the top authors earning rewards more than \$1000.
The median number of followers grows steadily without showing a flat period or a surge. 
Next, in Figure~\ref{sec5_1_2}, we investigate correlation between number of posts created by an author and rewards received by an author.
Again, we find that the line of average keeps itself above the line of median, indicating that some users post more blogs than average but do not receive more rewards as they may expect.
This may suggest that some authors slow down their posting speed while improving the quality of their posts.
Until \$30,000 rewards, the median number of posts shows a relatively stable growth.
After that, it experiences a fluctuation among the top authors.
Finally, in Figure~\ref{sec5_1_3}, we investigate correlation between VESTS owned by an author and rewards received by an author.
From the observation that the line of median lies nearly always below the line of average, we may infer that some users own higher shares of \textit{Steemit} than average but fail to increase their author rewards to the same scale.
By observing the line of median, we find that the median number of VESTS stays relatively flat at \$12.5 before author rewards reach \$10. 
The \$12.5 worth of VESTS is equal to 5 \textit{STEEM}, which is the minimum line that \textit{Steemit} suggests that its user transfer to \textit{VESTS} for the purpose of maintaining the user's account in a normal state.
From \$10 worth of author rewards, the median points start to disperse. It is interesting to find that many authors earning rewards more than \$10 maintain very low \textit{VESTS}, even lower than the suggested minimum line.
However, beyond \$10 worth of author rewards, the line of median still shows a positive trend.

Overall, our results illustrate a positive correlation between all the three factors and author rewards.
Our results do not imply causation, but they suggest that users, both authors and curators, consider the three factors as indicators of the potential popularity of posts.

\noindent \textbf{Curation rewards. }
As we have analyzed in section~\ref{s5_1}, the curation reward assigned to a single curator is affected by two factors. The first factor suggests that reward-driven curators cast their votes early and vote for posts with high potential for popularity. However, the second factor penalizes the early votes.
Regarding the potential popularity of posts, a reward-driven curator may leverage the number of followers/posts/VESTS of post authors as indicators.
However, due to the conflict suggestions from the two factors in the curation reward computation equation, it is hard to determine the best voting time for maximizing curation rewards.
Therefore, we decide to investigate what are the choices made by users in the data set.
To observe decisions made by different levels of shareholders, we divide users into four groups based on their owned \textit{VESTS}, namely low ($<10^4$), medium ($10^4-10^6$), high ($10^6-10^8$) and top ($>10^8$).
In Figure~\ref{sec5_2}, we plot the distribution of the accumulated $rshares$ that have been contributed by all votes to all posts every second after the post creation time.
Since $rshares$ is the most important factor of post rewards and also the key value of votes, this figure helps revealing users' strategies on selecting voting time.
As can be seen from results, there are two peaks that happened.
The first peak is at the sixth second, which indicates that many curators choose to vote for a post after six seconds (two blocks) of the post creation time.
This may reflect that many users are still using the early vote strategy.
The second peak is more interesting. It happens at the 1803 second, namely the first block after the 30-minute time period penalized by the second factor of the curation reward computation equation. 
This reflects that some users do understand the punishment mechanism and are deliberately avoiding from being penalized.
We can also observe that the low-level shareholders tend to not pay special attention to the voting strategies while all the other three groups of shareholders tend to contribute more to the two peaks and are more likely to vote earlier besides the two peaks.
The results of the distribution of accumulated curation rewards along time are shown in Figure~\ref{sec5_3}, which show a clear effect of the punishment mechanism on the early votes. The results suggest that curators who want to earn more rewards vote around the time of the second peak, namely the end of the 30-minute penalty period.

\subsection{Misuse of reward system}
The reward system in \textit{Steemit} may be a good way to reward contributors to the platform, but it may also be misused by some users in ways that deviate from the original intended goal in \textit{Steemit}, such as buying votes from bots to promote some meaningless posts for the purpose of earning profit.
In this subsection, we aim at understanding to what extent such behaviors are performed in \textit{Steemit}.

\noindent \textbf{Temporal correlation. }
We study this topic by investigating the temporal correlation between \textit{transfer} operations and \textit{vote} operations.
Concretely, if there is a suspicious vote cast by a curator to an author through a \textit{vote} operation, there should also be a suspicious fund transferred to the curator from the author through a \textit{transfer} operation that happens before the \textit{vote} operation is performed by the curator.
Specifically, we consider a \textit{transfer} operation is suspicious if the `memo' area (allowing sender to leave a message) of the \textit{transfer} operation only contains a link pointing to a recent post created by the sender.
If in addition, the recipient of the \textit{transfer} operation, after receiving the fund from the sender, votes for the post matching that link within the 7-days time window after the post creation time, we consider it as a suspicious trade between a post author and a voting bot.

After parsing the 18,172,530 \textit{transfer} operations in blockchain, we find that 5,031,737 (27.69\%) of them only contain a single post link in the `memo' area. By further investigating the \textit{vote} operation, we find that 2,939,051 (16.17\%) of the total \textit{transfer} operations are followed by a suspicious \textit{vote} operation voting for the post link within the 7-days time window, which are performed by the recipients of the funds.
Among all the fund recipients, the top one has performed such trade for 113,068 times and earned nearly 1 million USD.

\noindent \textbf{Voting time of suspicious curators.}
To further investigate the features of the curators suspected to be bots, in Figure~\ref{sec5_4}, we plot the cumulative distribution function (CDF) of the time gap between the post creation time and voting time. The solid line refers to the CDF of votes cast by all curators who ever received any amount of fund from the authors of voted posts at any time point in the past.
The dotted line describes the CDF of votes cast by curators suspected to be bots.
As can be seen, the curators suspected to be bots cast their votes much earlier than the average.
The dotted line reaches 98.9\% at the 604,800 second, which refers that nearly all their votes are sent within the 7-day time window and they are internationally contributing $rshares$ of their votes to the posts. 

\begin{figure}
\centering
{
    \includegraphics[width=7cm,height=6cm]{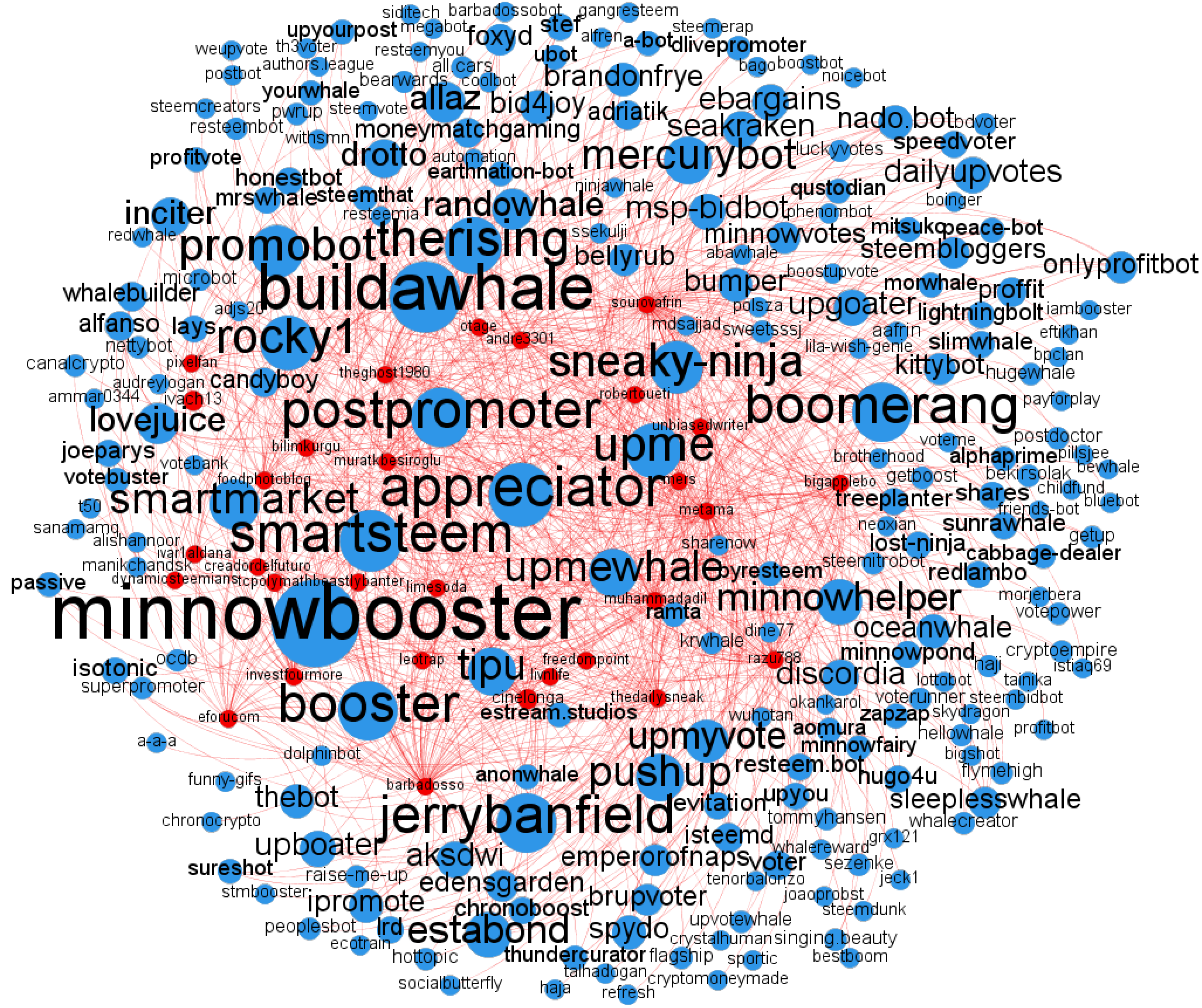}
}
\caption {\small Graph of \textit{transfer} operations sent from selected post authors (red) to the curators suspected to be bots (blue). The graph was plotted using Gephi~\cite{bastian2009gephi}.}
\label{sec5_5} 
\end{figure}

\noindent \textbf{An example bot network. }
In Figure~\ref{sec5_5}, we plot an example bot network, which describes the \textit{transfer} operations sent to the curators suspected to be bots from the top-30 post authors, who most frequently had contacted with these suspected curators.
The node sizes are weighted by weighted in-degree.
As can be seen, the thirty authors contacted a large number of suspected curators because they may need $rshares$ from more than one suspected curator to promote their posts.
We can also observe that most suspected curators have their account name containing suspected keywords, such as \textit{boost}, \textit{promote}, \textit{whale} and \textit{upvote}.

\begin{figure}
\centering
{
    \includegraphics[width=6cm,height=6.5cm]{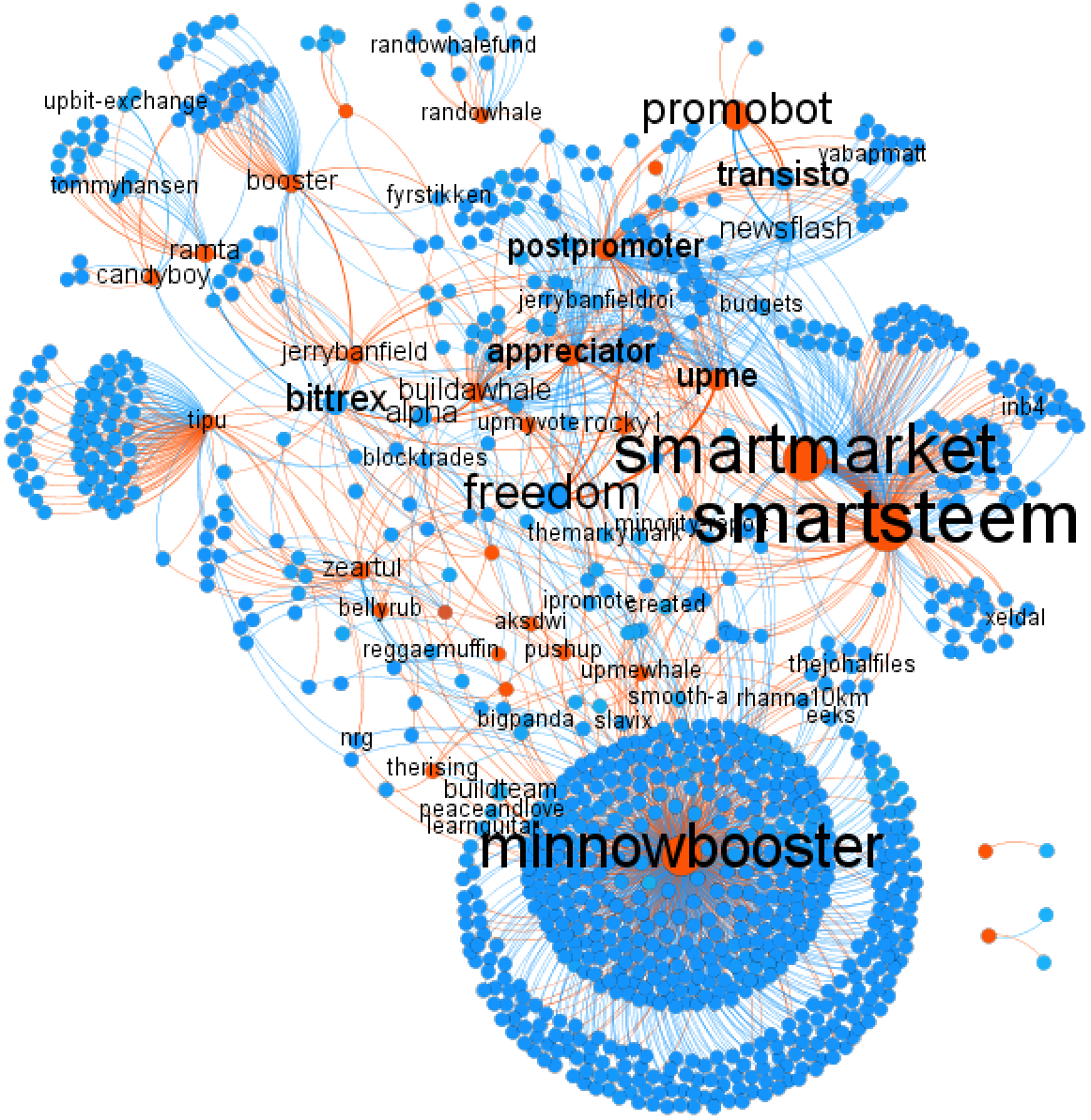}
}
\caption {\small Graph of \textit{delegate\_vesting\_shares} operations and \textit{transfer} operations among the top-30 curators suspected to be bots (red) and their suspected suppliers (blue). The graph was plotted using Gephi~\cite{bastian2009gephi}.}
\label{sec5_6} 
\vspace{-3mm}
\end{figure}

\noindent \textbf{An example supply network for bots. }
Finally, we analyze where do these suspected curators get large amounts of \textit{VESTS} to make their votes weighted by higher $rshares$.
After a deep analysis, we find that there is an underlying supply network, where some big shareholders delegate their \textit{VESTS} to the suspected curators through \textit{delegate\_vesting\_shares} operations and the suspected curators periodically pay `rent' to these \textit{VESTS} suppliers through \textit{transfer} operations. In Figure~\ref{sec5_6}, we plot an example supply network for bots, which describes the \textit{delegate\_vesting\_shares} operations sent to the top-30 suspected curators with highest earnings and the \textit{transfer} operations sent from the top-30 suspected curators with amount higher than \$100.
The node size reflects weighted degree and the edge thickness is weighted by the amount of transferred value.
From the graph, we find that most of the top-30 suspected curators own a number of suppliers and a few of them have many.
It is interesting to find that the top-1 supplier is the same user who also stands behind the top-1 witness presented in section~\ref{s4_2}.
We find that this user has earned more than 1.5 million USD by delegating his or her \textit{VESTS} to the curators suspected to be bots.

\subsection{Discussion}
In summary, we have discussed the design principle, current use status and underlying misuse of the cryptocurrency-driven reward system of \textit{Steemit} in this section.
The rise of social bots and the harm caused by them to the online ecosystems has been widely recognized~\cite{ferrara2016rise}.
Our results in this section reveal that bots have broadly appeared in the emerging incentivized blockchain-based social media networks.
In \textit{Steemit}, the monetary value of cryptocurrencies and the deep integration of social operations and value-transfer operations have jointly led to the special supply chain, where bots buy voting powers from big stakeholders and sell the voting powers to users who want to promote posts.
Clearly, such behaviours violate the original intention of the reward system and may eventually result in a front page occupied by valueless posts or spams promoted by bots.
Moreover, due to the decentralized operation, it is difficult to delete these problem posts appeared at the front page because no single entity has this power.
Our discussion in this section also suggests a way of detecting bots through investigating the temporal correlation between \textit{transfer} operations and \textit{vote} operations. 
We believe this detection approach is effective at the current stage when bots prefer to trade with the native cryptocurrency $STEEM$, which is detectable through the Steem-blockchain. 
However, smarter bots may hide the temporal correlation between \textit{transfer} operations and \textit{vote} operations by asking their clients to pay them through other cryptocurrencies such as Bitcoin or even privacy-preserving cryptocurrencies such as Zerocash~\cite{sasson2014zerocash}.
Then we need to rely on other approaches, such as Deep Neural Networks~\cite{kudugunta2018deep}, to detect bots.

\section{Related work}

In recent years, due to the rapid growth and consistent popularity, social media platforms have received significant attention from researchers. 
A large number of research papers have analyzed several popular social media platforms from various perspectives~\cite{stoddard2015popularity,wang2013wisdom,de2014mental,yang2018understanding,singer2016evidence,glenski2017consumers,tan2015all,hessel2016science,anderson2012discovering}.
Tan \textit{et al.}~\cite{tan2015all} investigated users' behavior in \textit{Reddit} and found that users continually post in new communities.
Singer \textit{et al.}~\cite{singer2016evidence} observed a general quality drop of comments made by users during activity sessions.
Hessel \textit{et al.}~\cite{hessel2016science} investigated the interactions between highly related communities and found that users engaged in a newer community tend to be more active in their original community.
In~\cite{glenski2017consumers}, the authors studied the browsing and voting behavior of \textit{Reddit} users and found that most users do not read the article that they vote on.
An extensive survey of recent research on \textit{Reddit} is provided in~\cite{medvedev2018anatomy}.
Besides \textit{Reddit}, several other social media platforms have also been analyzed by researchers.
Wang \textit{et al.}~\cite{wang2013wisdom} analyzed the \textit{Quora} platform and found that the quality of Quora’s knowledge base is mainly contributed by its user heterogeneity and question graphs.
Anderson \textit{et al.}~\cite{anderson2012discovering} investigated the \textit{Stack Overflow} platform and observed significant assortativity in the reputations of co-answerers, relationships
between reputation and answer speed.

Recent advances in blockchain and distributed ledger technologies in terms of scalability~\cite{kokoris2018omniledger,kalodner2018arbitrum}, efficiency~\cite{tomescu2017catena,chase2016transparency} and privacy~\cite{sasson2014zerocash,kosba2016hawk,camenisch2017practical} have empowered blockchains to support various services beyond money transfer, including  incentivized blockchain-based social media platforms such as \textit{Steemit}.
Recently, this new type of social media platform has drawn some attention from researchers.
Thelwall \textit{et al.}~\cite{thelwall2017can} analyzed the first posts made by 925,092 \textit{Steemit} users to understand the factors that may drive the post authors higher rewards.
Their results suggest that new users of \textit{Steemit} start from a friendly introduction about themselves rather than immediately providing useful content.
In a very recent work, Kiayias \textit{et al.}~\cite{kiayias2018puff} studied the decentralized content curation mechanism from a computational perspective. They defined an abstract model of a post-voting system, along with a particularization inspired by \textit{Steemit}.
Through simulation of voting procedure under various conditions, their work identified the conditions under which \textit{Steemit} can successfully curate arbitrary lists of posts and also revealed the fact that selfish participant behavior may hurt curation quality.
To the best of our knowledge, the work presented in this paper is the first research that investigates the reward system and decentralization features of incentivized blockchain-based social media platforms through a rigorous empirical analysis of the operations reflected in the underlying blockchain data.

\section{Conclusion}
In this paper, we presented an empirical analysis of \textit{Steemit}, a blockchain-based incentivized social media platform where no single entity can take control of the information and users are rewarded for the contributions they make. 
We analyzed over 539 million operations performed by 1.12 million users during the period 2016/03 to 2018/08. 
Our results show interesting details about two core features of \textit{Steemit}, namely its decentralized management and its reward system. 
Our study on decentralization in \textit{Steemit} shows the actual level of decentralization in \textit{Steemit} is far lower than the ideal level, indicating that DPoS consensus protocol may not be a desirable approach for establishing a highly decentralized social media platform.
Our analysis of the reward system reveals the fact that more than 16\% transfers of cryptocurrency in \textit{Steemit} are sent to curators suspected to be bots and also finds the existence of an underlying supply network for the bots, both suggesting that the current cryptocurrency-driven reward system in \textit{Steemit} is under severe misuse that deviates from the original intended goal of rewarding high-quality contents.. 
Overall, we believe that the results in this paper provide insights on the current state of the emerging blockchain-based social media platforms including the effectiveness of the design and the operation of the consensus protocols and the reward system.

\renewcommand\refname{Reference}

\bibliographystyle{plain}
\urlstyle{same}

\bibliography{main.bib}

\end{document}